# Island instability as a mechanism of destabilization and collapse of current sheets

Shaikhislamov I.F.

Dep. of Laser Plasmas, Institute of Laser Physics, Novosibirsk, 630090, Russia

e-mail: ildar@plasma.nsk.ru

**Abstract.** The work shows that a single elongated island immersed in the quasi-neutral current sheet makes it MHD unstable. Typical values of growth rate are found to be several percent of inverse Alfven time for broad sheets. Hall dynamics greatly enhance instability and growth rate reaches one-tenth of ion cyclotron frequency when sheet width is comparable to ion inertia length. A week square-root dependence of increment on island width and length is derived both from numerical simulation and analytical analysis. At the non-linear phase of evolution a phenomenon of fast impulsive intensification of current is found. After finite time of about ten ion-cyclotron periods it ends by a collapse of the sheet. It is shown that Hall dynamics is responsible for such a behavior. Possible implications for magnetotail substorms are discussed.

## 1. Introduction

The neutral line model of substorms in near-Earth space [1] and in other astrophysical environments provides a coherent picture of how the stored magnetic energy in the magnetotail is converted to the kinetic and thermal plasma energy. Reconnection of the opposite magnetic field lines is thought to be responsible for the fast and efficient energy release [2-4]. However, the detailed understanding of how a new neutral line forms within the preexisting closed field lines of the plasma sheet, and how the current sheet becomes sufficiently thin to start the reconnection remains a challenging and fundamental issue. Space observations at near-Earth distances [5, 6] reveal that between a period of sluggish growth (~0.5-1.5 h) and the onset of the substorm expansion phase, a fast explosive like intensification of the thin tail current sheet occurs (~1 min) followed by a sudden disruption on a very short time scale (<<1 min). These show that there are at least three different time scales and a viable theory of substorms is expected to account for all of them.

Despite a large number of processes proposed for the triggering of substorms, the problem remains a challenge that motivates the search for new mechanisms. A collisionless tearing instability, being for a long time an obvious candidate for such a role, has been shown to be stabilized [7] by the normal component of magnetic field $B_z$. The hypothesis of anomalous resistivity generated by various current driven micro-instabilities is not supported by experimental observations [8]. One of the most developed models is the forced reconnection in which an external electric field drives plasma toward the neutral line. For this a region of finite resistivity is also needed, though it was shown that Hall dynamics makes the reconnection rate insensitive to the actual value of the resistivity [9]. Given sufficient time, current sheet driven by external field gradually thins and intensifies. When its width becomes well below ion-inertia length the current density exhibits fast and impulsive enhancement [10].

In this work we investigate altogether different mechanism of current sheet destabilization and intensification. It is essentially internal and is driven by the excessive magnetic energy of the sheet, but, unlike other scenarios, it doesn't need any resistivity. It doesn't rely on kinetic effects or finite electron mass as well. The process operates in an island configuration of magnetic field studied before in the context of island coalescence. However, if the chain of islands could be envisioned only as a product of preceding tearing instability and multiple-point reconnection, a single and elongated island of small amplitude embedded in the sheet could be considered as an independent and a-priory configuration. A several new features will be shown in the paper. Numerical simulation (Sec.3) and analytical analysis (Sec.4) demonstrate that MHD instability successfully develops even if only one island is present in the sheet. For the sheets with width comparable to the ion-inertia length a growth rate is strongly enhanced by the Hall (electron) dynamics. It depends rather weakly on the island parameters (island width and length) and could reach quite large values $\sim 0.1 \cdot \omega_{ci}$. At the non-linear phase of evolution (Sec.5) a striking phenomena is reported – fast intensification and collapse of the sheet, when current density behaviors like $J \sim (1-t/\tau)^{-1}$ and goes to infinity for a finite time ($\tau \sim 10/\omega_{ci}$).

## 2. Geometry and Hall MHD equations

In this work a collisionless MHD with a generalized Ohm's law will be used:

$$E + \frac{V \times B}{c} = \frac{J \times B}{nec} - \frac{\nabla P_e}{ne}. \tag{1}$$

The right side of (1) consists of the so-called Hall terms with electron mass ignored. Throughout the paper the Geomagnetic tail coordinate system will be used. The problem is restricted to two dimensions with $\partial/\partial y = 0$; $x$-axis is directed along the magnetotail. Choosing for characteristic values the size of the current sheet $d$ in $z$-direction, magnetic field $B_o$, particle density $n_o$, corresponding Alfven speed $V_A$ and pressure $B_o^2/4\pi$ we arrive to the following dimensionless Hall MHD equations:

$$\frac{\partial n}{\partial t} + \nabla \cdot (nV) = 0,$$

$$\frac{\partial P}{\partial t} + \nabla \cdot (PV) + (\gamma - 1) P \nabla \cdot V = 0,$$

$$n\left(\frac{\partial V}{\partial t} + (V\nabla)V\right) = J \times B - \nabla P, \tag{2}$$

$$\frac{\partial B}{\partial t} = \nabla \times (V \times B) - \chi \cdot \nabla \times \left(\frac{J \times B - \nabla P_e}{n}\right),$$

$$J = \nabla \times B.$$

Characteristic time of the problem is Alfven time $\tau_A = d/V_A$. Parameter $\chi = c/(d \cdot \omega_{pi})$ defines a domain where Hall effects are important – at scales comparable or smaller than ion inertia length ($d \leq \lambda_i$). Initially plasma is at rest and in equilibrium. To demonstrate the processes under consideration more clearly, we simplify equations further by ignoring the second part of the Hall term $\nabla \times [(\nabla P_e)/n]$, which corresponds to either isothermal or cold electrons. For the perturbed values $v, b, p$ linearized equations are:

$$\frac{\partial p}{\partial t} + \nabla \cdot (vP) + (\gamma - 1)P\nabla \cdot v = 0,$$

$$n\frac{\partial v}{\partial t} = J \times b - B \times \nabla \times b - \nabla p, \qquad (3)$$

$$\frac{\partial b}{\partial t} = \nabla \times (v \times B) - \chi \cdot \nabla \times \left(\frac{J \times b - B \times \nabla \times b}{n}\right).$$

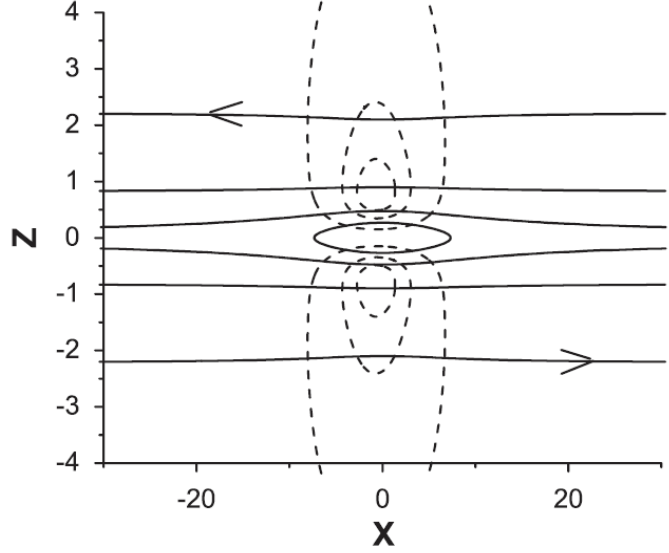

**Figure 1.** Magnetic field lines of equilibrium configuration (full curves). Broken curves show trajectories of plasma motion generated by instability.

Capital letters are reserved for unperturbed values. For the model of quasi-neutral current sheet we use here a well known general class of two-dimension equilibria [11]: $\nabla^2 A_y = (B_o/d) \cdot exp(-2A_y/B_o d)$, where $A_y$ is a flux function. Its approximate solution suitable for the extended magnetotail is:

$$A_y(x,z) = B_o d \cdot ln[\cosh(F z/d)/F], \qquad (4)$$

where $F(x)$ is an arbitrary function slowly varying along $x$ over distances much larger than the sheet half-width $d$. The value of normal component of magnetic field on the neutral line is given by $B_z = -B_o(d \cdot F'/F)$. In case of $F \equiv 1$ a Harris sheet is retained. For the sheets with non-zero background density, the equilibrium density is given by $n = n_o F^2 / \cosh^2(F z/d) + n_\infty$, while thermal pressure in dimensionless units is expressed as: $P = (\beta/2) \cdot n$, $\beta = 8\pi n_o (T_i + T_e)/B_o^2$.

Island is an O-point with closed magnetic field lines around it. It is described by function $F(x)$ having a local maximum (while X-point by a local minimum). Below a following model function will be used:

$$F = 1 - \varepsilon + \frac{\varepsilon}{1 + x^2/L^2}. \tag{5}$$

It describes O-point located at $x=0$ embedded in a one-dimensional Harris sheet, as shown in Fig.1. The width of the island at small value of parameter $\varepsilon$ is $4\sqrt{\varepsilon}$, while its length in $x$-direction is given by the parameter $L$. Maximum value of the reversing $B_x$ component is equal to $B_o$ (at the $x=0$ section), while maximum of $B_z$ component on the neutral line is $|B_z|_{max} \approx 9/(8\sqrt{3}) \cdot \varepsilon/L$. Results with other model functions will be mentioned as well. It is worth noting that the only equilibrium studied so far is given by the flux function $A_y(x,z) = B_o d \cdot \ln[\cosh(z/d) + \varepsilon \cos(x/d)]$, that is a periodic configuration of multiple islands separated by X-points. In this case the length of islands is equal to the sheet width.

Equations (2) have been solved numerically on a non-uniform rectangular mesh. Code implementation is similar to one, described in [12]. The simulations were employed typically in the box $0 \leq z \leq 10$, $-70 \leq x \leq 70$ with a number of grid points up to $N_z$, $N_x = 400$. In some runs the resolution in $z$-direction at the center of the current sheet was better than $0.005 \cdot d$. Fix conditions were chosen at the boundaries, where variation of all values relative to the equilibrium was put to zero.

## 3. Linear phase of island instability

The process we study here has been investigated previously in the context of island coalescence [13] and was called correspondingly coalescence instability. It was established that a chain of islands is subjected to ideal MHD instability [14, 15] with relatively large increment (compared to the collisionless tearing) which is practically independent on plasma $\beta$ [16]. Plasma flows along $x$-direction inside each island with maximum velocity at O-point and back outside the island. Generally it is thought that the source of instability is the attraction of parallel current filaments (O-points). However, it is easy to show that island coalescence is the aftereffect, but not the reason of instability. Rather, it is the intrinsic property of each island independently. Let's consider two islands symmetrically placed at $x$-axis. Linearized MHD equations allow transformation $\boldsymbol{v} \to -\boldsymbol{v}; \boldsymbol{b} \to -\boldsymbol{b}; p = -p$. Thus, if $\boldsymbol{v}(x,z)$ is such unstable solution that islands are drawn to each other, then there exists identically unstable solution $-\boldsymbol{v}(x,z)$ with islands being moved apart from each other. The reason why coalescence always occurs in the periodic chain of islands (and this is the only configuration studied so far) is obvious – neighboring islands are drawn to or moves apart in pairs, and any island collides either with the left or with the right neighbor. Because of the given arguments further on we shall call this process island instability.

Numerical simulation confirms that instability under consideration successfully develops for a single island as well. For the sake of simplicity, at first results of MHD simulations without Hall term ($\chi=0$) are presented. Plasma motion generated by instability is demonstrated in Fig.1. The flow goes parallel to $x$-axis inside the island, sharply turns at some $|x|<L$ and returns over quite wide region outside it. For the random initial perturbation the flow inside the island has equal probability to develop in either of directions. The dependence of growth rate on geometric parameters $\varepsilon$ and $L$ is shown in Fig.2. It suggests that a combination $\varepsilon/L \approx B_z$ is important. Mean-square fit gives approximate empirical formulae $\gamma \tau_A \approx 0.4 \cdot (\varepsilon/L)^{2/3}$.

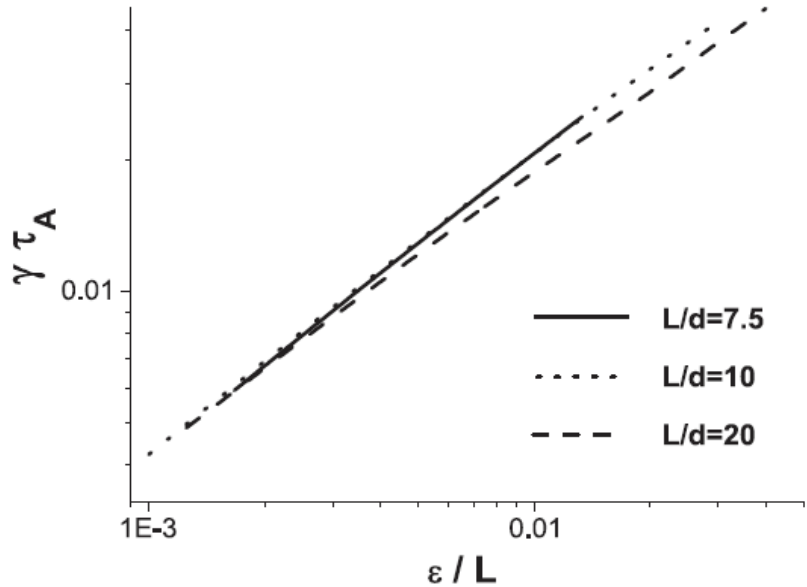

**Figure 2.** Dependence of growth rate on island parameters $\varepsilon$ and $L$ in the ideal MHD limit $\chi \to 0$. Results are presented in dependence on a combination $\varepsilon/L$ and grouped by the value of $L$.

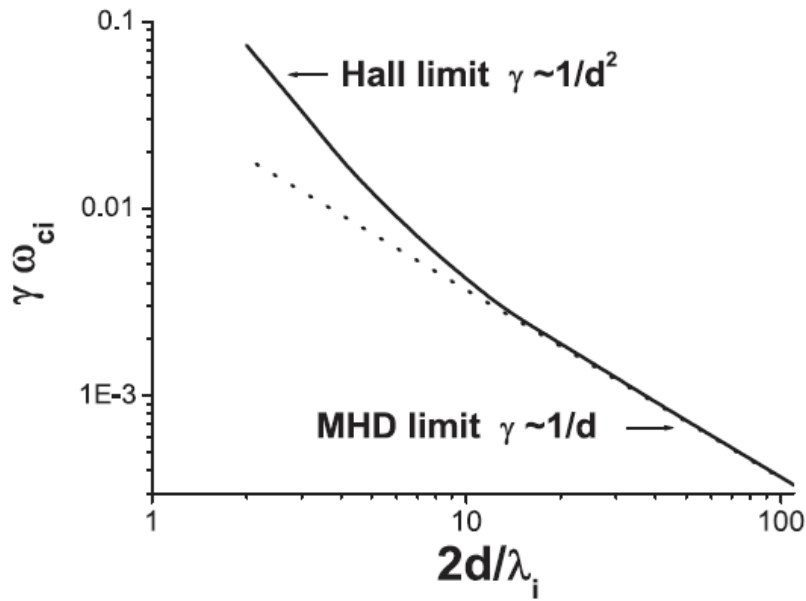

**Figure 3.** Dependence of growth rate on the width of the current sheet. The broken line shows the ideal MHD limit.

Typical values of the increment at $B_z > 0.01$ are several percent of inverse Alfven time. Plasma beta for this calculations was $\beta = 4$. The effect of compressibility was found to be negligible - less than *3%* over wide range $1 \leq \beta \leq 10$. It was found that increment doesn't depend very much as well on the type of model function $F(x)$. In general, it is larger for a more compact and localized islands. For example, for $F = 1 - \varepsilon + \varepsilon/(1 + x^6/L^6)$ and $F = 1 - \varepsilon + \varepsilon \cdot exp(-x^2/L^2)$ the growth rate is approximately *15%* greater than for the function given by (5).

    Hall dynamics significantly complicates the picture, mainly because it couples shear Alfven wave ($v_y$ component of plasma velocity) to the in-plane motion [17]. On the other hand, the in-plane Hall current, like in the tearing instability [18], is qualitatively similar to the in-plane ion motion. There is a simple approximate relation between them: $\boldsymbol{j} \approx -\chi \nabla \times \nabla \times \boldsymbol{v}$. The convection of magnetic field by this current adds to the convection by plasma motion and, thus, increases island instability. This has been found in other works where coalescence process was studied in the frame of Hall MHD [12]. The only noticeable difference between Hall current and plasma velocity fields is that the first is significantly more localized around the neutral line.

    In Fig.3 the dependence of increment on the total width of the current sheet *2d* is shown. For this case an inverse of ion-cyclotron frequency $\omega_{ci}$ is more suitable as the characteristic time. As one can see, already at $\chi > 0.5$ ($2d/\lambda_i < 4$) the instability is dominated by Hall dynamics and the increment scales correspondingly as $\gamma \sim 1/d^2$, while for ideal MHD dynamics the scaling is $\gamma \sim 1/d$. For the sheets with width comparable to the ion inertia length the growth rate reaches $0.1 \cdot \omega_{ci}$ even at quite small values of the normal component $B_z \geq 0.01$. Because Hall dynamics strongly influences the island instability, further on the problem will be considered in the frame of Hall MHD. Dependence of the growth rate on geometric parameters $\varepsilon$ and $L$ for the sheet with $d = \lambda_i$ is presented in Fig.4. It suggests that in this case a combination $\varepsilon/L^2$ works. Data could be described by empirical formulae $\gamma \omega_{ci} \approx 0.43 \cdot (\varepsilon/L^2)^p$ with $p \approx 0.23$.

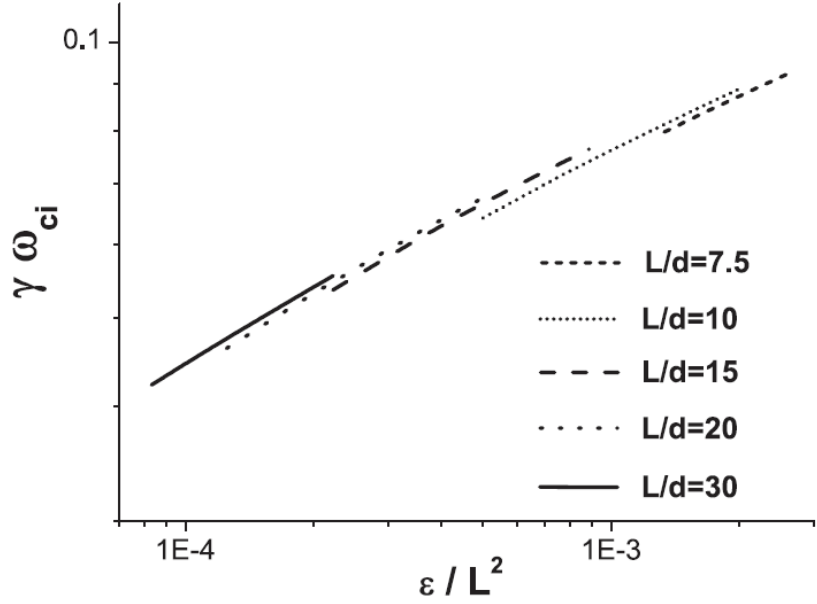

**Figure 4.** Dependence of growth rate on island parameters $\varepsilon$ and $L$ at $d = \lambda_i$. Results are presented in dependence on a combination $\varepsilon/L^2$ and grouped by the value of $L$.

## 4. Analytical analysis in the Hall limit

To verify and understand results of numerical simulation we perform analytical analysis of instability. As has been shown above, electron motion totally dominates over ion motion at $\chi \sim 1$. Thus, we consider pure Hall dynamics which is described by the equation of magnetic field evolution:

$$\frac{\partial \boldsymbol{b}}{\partial t} = -\nabla \times (\boldsymbol{J} \times \boldsymbol{b} - \boldsymbol{B} \times \nabla \times \boldsymbol{b}). \tag{6}$$

Here we have chosen ion inertia length $\lambda_i$ and a Hall time $(d/\lambda_i)^2/\omega_{ci}$ as the characteristic length and time of the problem. Also, for the sake of simplicity, density was taken to be constant. For the out of plane components it can be written:

$$\frac{\partial a_y}{\partial t} = -(\boldsymbol{B} \cdot \nabla) b_y$$

$$\frac{\partial b_y}{\partial t} = (\boldsymbol{B} \cdot \nabla) \nabla^2 a_y + (\boldsymbol{b} \cdot \nabla) \nabla^2 A_y \tag{7}$$

$$\frac{\partial^2 a_y}{\partial t^2} = -(\boldsymbol{B} \cdot \nabla)\left((\boldsymbol{B} \cdot \nabla) \nabla^2 a_y + (\boldsymbol{b} \cdot \nabla) \nabla^2 A_y\right),$$

were $\boldsymbol{b} = -\boldsymbol{e}_y \times \nabla a_y$. To make this essentially two-dimensional problem tractable, we ignore details of solution dependence on $x$ coordinate and take it in the form $a_y \sim exp(\gamma t) \cdot sin(kx) \cdot a_y(z)$ with some unspecified $k<<1$. Accordingly, for the unperturbed values we keep dependence on $x$ only for the odd functions of $x$. Obviously, this approximation may be applied only at $x<<L$. Note that $k$ is meant here to be a parameter by which solution is correctly described near the origin rather than variable. Its actual value could be found only by 2-D treatment. Simulation shows that it is a complex function of $L$ and that $kL>1$. Despite this uncertainty, it appears that most important features of instability could be derived nonetheless.

Equations, thus reduced to one dimension, could be treated by the method of a singular layer. For the initial configuration we take a model step-like current sheet with the main component of magnetic field $B_x = 1$ at $z \geq 1$ and $B_x = z$ at $z<1$. Near the neutral line, where main field vanishes, we take into account a small normal component $B_z = -\alpha x/L$. Parameter $\alpha << 1$ is in our case the expansion parameter. Solution outside singular layer $z > \delta$ is well known from the classic analysis of the tearing mode. It is $a_y \approx exp(-kz)$ at $z \geq 1$ and $a_y \approx z/k$ at small $z$. The jump across origin is $\Delta' = 2/k$. Inside the layer $\delta$, where only normal component operates, the last equation of (7) is reduced to

$$\gamma^2 a_y = -(B_z \partial/\partial z)(B_x \partial/\partial x)\nabla^2 a_y \approx \alpha/L \cdot \partial^2 a_y/\partial z^2. \qquad (8)$$

Here we used an approximate treatment of $x$-dependence by making substitution: $kx \cdot cos(kx) \approx sin(kx)$. Instability is driven by magnetic term $-(\partial B_z/\partial x)(\partial B_x/\partial z) = (\partial^2 A_y/\partial x^2) \cdot (\partial^2 A_y/\partial z^2)$ which is positive for O-point and negative for X-point. This is why it works only for the island configuration. From (8) it follows that the jump across the layer $2\delta$ is $\Delta' = 2\gamma^2 \delta L/\alpha$, and the increment $\gamma^2 = \alpha/(\delta kL)$. To find the width of the layer $\delta$ we use second equation of (7):

$$\gamma a_y = -z \cdot k \cdot b_y + (\alpha/kL) \cdot \partial b_y/\partial z. \qquad (9)$$

From (9) it immediately follows that $\delta \approx \sqrt{\alpha/L}/k$. Finally, increment follows as $\gamma \approx (\alpha/L)^{1/4}$, which, luckily, is independent of *k*. To compare it with the numerical results we have to substitute $\alpha$ by $2\varepsilon/L$. A more elaborate treatment of one-D equations for the Harris sheet rather than step-like profile yields an expression $\gamma \approx 0.52 \cdot (\varepsilon/L^2)^p$ with $p\approx0.23$. This result, despite all approximations made, catches a right parametric dependence, and quantitatively differs from the numerical data only by ~20%. The found dependence could be given an elegant geometrical meaning. An island' aspect ratio (length divided by width) is equal to $A = L/(2\sqrt{\varepsilon})$ and, thus, $\gamma \sim 1/\sqrt{A}$.

Note, that we haven't in fact found relevant solution of (7) because the value of appropriate wavenumber *k* is not known. However, it gives right qualitative picture. In Fig.5 the structure of perturbed values at *x=0* section obtained from the numerical simulation are shown. A singular layer is clearly seen in a distribution of the Hall current $j_x$. Performed analysis suggests that instability in the Hall limit is generated at close vicinity of the O-point and is determined by a variation of the normal component across it $\partial B_z/\partial x \sim \varepsilon/L^2$ rather than its absolute value. Behavior of solution in the far region $x \sim L$ adjusts itself correspondingly to the structure near the origin. Namely this property validates the used approximations, which are not strictly justified otherwise.

In the opposite limit (ideal MHD without Hall term) the structure of instability is more complex. Equations (3) at the origin point $x,z \to 0$ have only one non-zero term: $n \cdot \partial^2 v_x/\partial t^2 = \partial/\partial x(P \cdot \nabla \cdot v)$. Thus, if plasma is totally uncompressible ($\nabla \cdot v = 0$), the configuration is stable. However, numerical simulation reveals that plasma experiences compression and rarefaction in the regions of sharp turning of flow. This shifts the balance towards instability. Instability is generated close to the regions of maximum of the normal component $B_z$. This is why the increment in this case depends on the geometric parameters in a combination $\varepsilon/L \sim B_{z,max}$.

In both cases a general source of instability is the excess of magnetic energy of the current sheet. A particular reason of its existence in the island configuration is non-zero $B_z$ component of magnetic field that allows self-consistent motion of ions and electrons at the neutral line. This is in contrast to the tearing instability, which is strongly stabilized by the normal component.

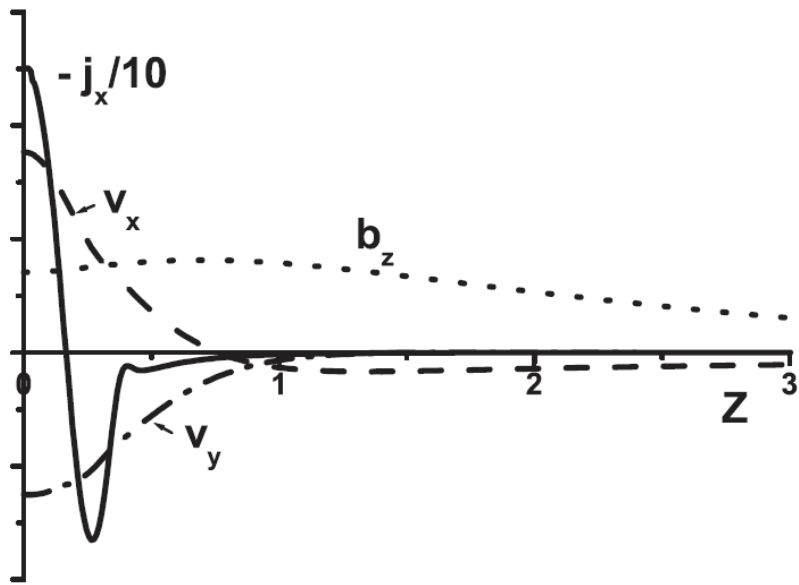

**Figure 5.** Distribution of perturbed values in section $x = 0$ calculated for the parameters $d = \lambda_i$, $L/d = 10$ and $\varepsilon = 0.1$.

## 5. Non-linear phase

When distortion of initial configuration becomes significant, development enters non-linear phase which is characterized by a strong convection of magnetic field. Subsequent evolution reveals a new striking feature – a collapse of the current sheet. Maximum current density increases and its half-width decreases with faster and faster rate as demonstrated in Fig.6. In this figure time evolution of various perturbed values is shown for the initial configuration $d = \lambda_i$, $L/d=10$, $\varepsilon = 0.1$. One can clearly see linear growth phase ($t\omega_{ci} < 160$), saturation of instability and still further and slower increase ($160 < t\omega_{ci} < 230$). When perturbation of the current density becomes comparable to the initial one the sheet experiences altogether different and dramatic behavior. Main current $J_y$ and $x$- component of the Hall current start to rise with much faster and ever increasing rate ($230 < t\omega_{ci} < 250$).

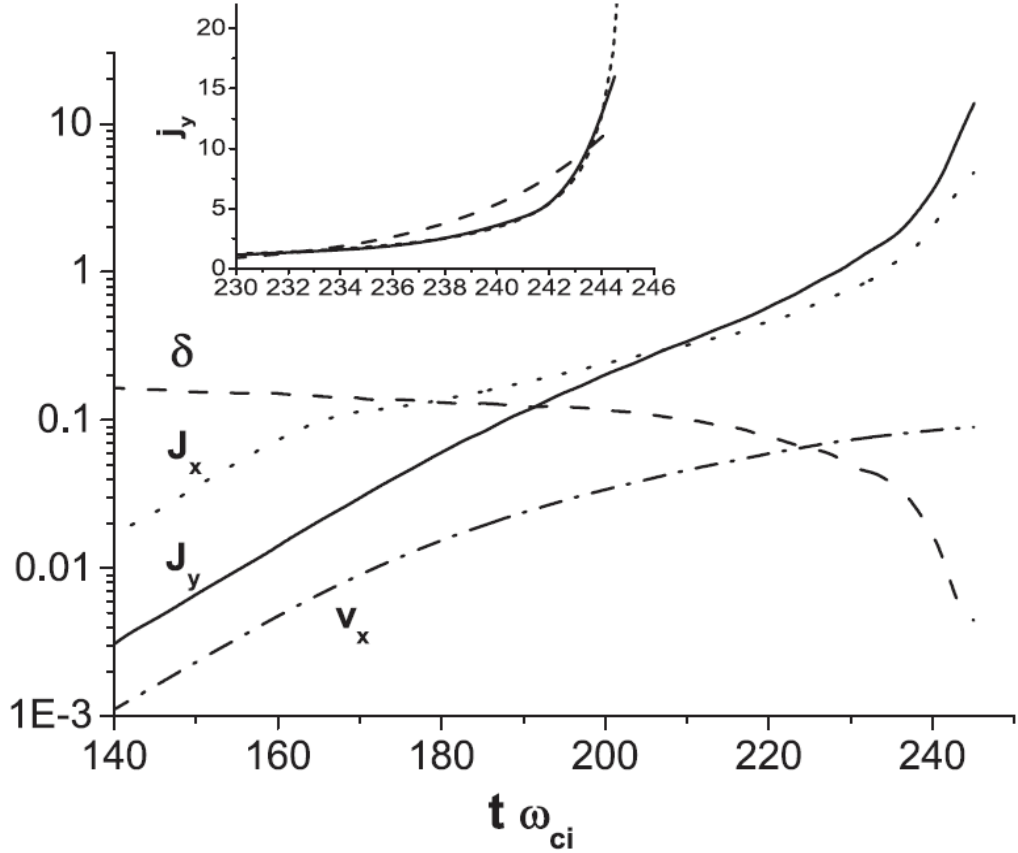

**Figure 6.** Non-linear evolution of main current density $j_y$, its half-width $\delta$, Hall current $j_x$ and plasma velocity $v_x$. Initial configuration is given by $d = \lambda_i$, $L/d = 10$ and $\varepsilon = 0.1$. Inserted panel: current density $j_y$ during collapse phase. The broken curve shows best exponential fit, while short-broken—fit by function $j_y = [1 - (t\omega_{ci} - 227)/19]^{-1}$.

A closer look reveals that there is no piling up of flux near the neutral line and magnetic field profile across the sheet remains monotonic. In fact, a sub-sheet is created with a relatively small jump of magnetic field across it $\Delta B_x << 1$. However, the width of this sub-sheet becomes exceedingly small, while current density exceedingly large. Their product $J_y \cdot \delta$ remains constant during the collapse. The length of the thin layer along x slightly decreases with time, but its aspect ratio increases dramatically. The increase of $J_y$ takes place in the region where the convection flow $v - j/ne$ converges. Because of this, the normal magnetic component inside it strongly decreases. Thus, in the end the sub-sheet resembles a very thin, stretched and intense one-dimensional structure superimposed on the initial and practically unchanged current sheet. In the region where flow diverges the field is dragged away from the neutral line, current density decreases and a magnetic bubble is formed. All this features are demonstrated in Fig.7 which shows magnetic field lines and Hall current lines at the near collapse time.

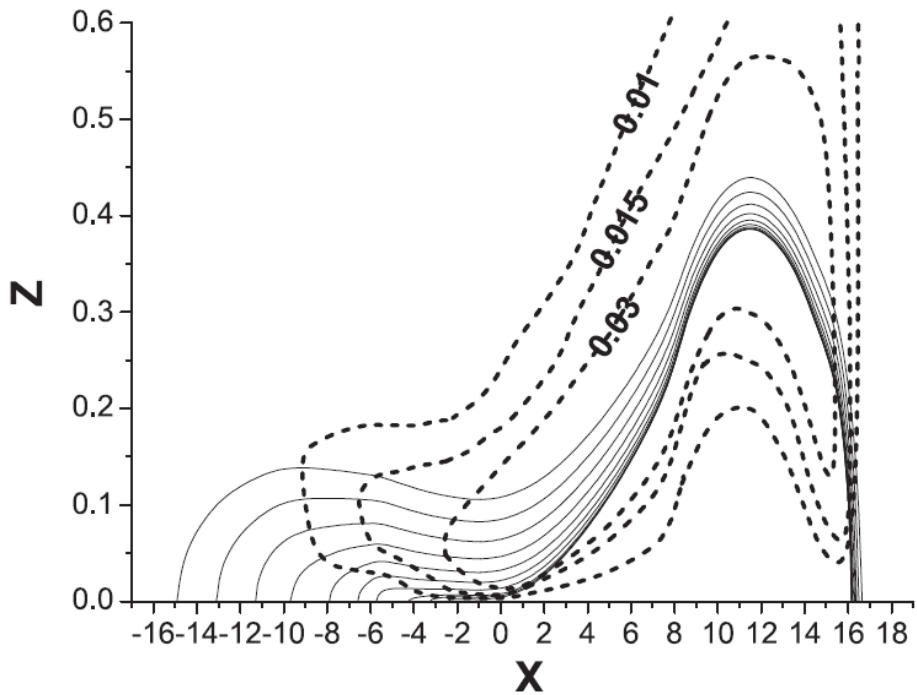

**Figure 7.** Contours of flux function $A_y$ (———) and $B_y$ component (- - - -) at $t\omega_{ci} = 244$. The position of the maximum of main current is $z = 0$; $x = -2.5$. Flux contours are numbered in such a way that at section $x = -2.5$ magnetic field is $B_x = 0.2 \cdot B_0$ for the farthest line and declines by $0.02 \cdot B_0$ for each subsequent one.

Hall dynamics, in fact, allows solutions that break for a finite time duration. At a region where electron velocity converges, the normal component of field $B_z$ decreases while current concentrates closer and closer to the neutral line. To describe this process approximately, we ignore $B_z$ altogether, as well as any derivative in respect to $x$ in comparison to $z$ ($\partial/\partial x \ll \partial/\partial z$). Then non-linear equation (6) in components is:

$$\partial A_y / \partial t = -B_x \cdot \partial B_y / \partial x$$

$$\partial B_y / \partial t = -B_x \cdot \partial J_y / \partial x \qquad (10)$$

Applying $\partial^2/\partial z^2$ and $\partial/\partial z$ to the first and second equation correspondingly and taking them at the neutral line $z=0$ where $B_x \to 0$ we obtain: $\partial J_y/\partial t = -2 J_y \cdot \partial J_x/\partial x$; $\partial J_x/\partial t = J_y \cdot \partial J_y/\partial x$. Combination of those two equations gives:

$$\frac{\partial}{\partial t}\frac{1}{J_y}\frac{\partial}{\partial t} J_y = -2\frac{\partial}{\partial x} J_y \frac{\partial}{\partial x} J_y \approx J_y^2/L^2 \qquad (11)$$

Here we ignored details of the dependence on *x*-coordinate assuming that maximum of $J_y$ current is generated at a converging point with maximum of $\partial J_x/\partial x$. There is a solution of (11) in the form

$$J_y(t) = 2 \cdot J_x(t) = J_o/(1 - t/\tau), \qquad (12)$$

with a characteristic time $\tau = L/J_o$. It shows that currents reach infinity for a finite time. This time depends on the amplitude at the start of the collapse, for which an estimate could be assumed $J_o \sim 1$. Results of simulation shows that during the collapse current behavior couldn't be fitted either with polynomial or exponential growth, while (12) provides a very good description (see inserted panel in Fig.6). For the presented case best fit gives the collapse time $\tau = 19/\omega_{ci}$. In the frame of used ideal Hall-MHD equations no mechanism to restrict current growth was found, except numerical limitation imposed by finite grid size.

In previous works on the island coalescence no such phenomena has been reported. There may be several reasons. First, the interaction between islands strongly affects the non-linear evolution and changes the flow pattern completely. When coalescence of islands takes place, convection velocity along *x* direction slows down and reverses [12]. Second, simulation and analysis show that collapsing structure is characterized by high aspect ratio ($\delta \ll L$), and that it develops at the part of the sheet where normal component becomes, in the course of prior evolution, very small. However, if the flow pattern generated by instability is not sufficiently stretched because of initial configuration, no such conditions could be reached. Note that in most studies of coalescence the equilibrium configuration used was characterized by $d \approx L$.

In the frame of ideal MHD without Hall term no collapse was observed. Instability saturates when plasma velocity reaches some fraction of Alfven speed (*~0.2-0.4*). After this, maximum velocity doesn't grow while the sheet is compressed further and current density continues to rise, albeit slower, due to the action of convective flow. However, even for broad but finite sheets, for example $d = 10 \cdot \lambda_i$, collapse phase is eventually reached. In this case preceding evolution takes longer time (in corresponding dimensionless units $d/V_A$) simply because the growth rate of instability is smaller. When the width of the perturbed current becomes sufficiently small, Hall dynamics starts to determine evolution, which ends in the collapse. Obviously, when intense structure thins down to electron inertia length, finite electron mass will restrict the current growth. However, the implications of this are beyond the scope of present paper.

## 6. Conclusion

In this work we studied stability of the current sheet in a single island configuration. It was found to be MHD unstable with the increment of order of $0.1 \cdot \omega_{ci}$ for the sheets with width of about ion-inertia length. Presented results of numerical simulation and approximate analytical analysis shows that, namely, a normal component of magnetic field associated with the O-point structure provides a way for the release of excessive energy because allows a self-consistent plasma motion at the neutral line. It was argued that this instability, named previously in the context of island coalescence, develops in fact independently of this process. In the non-linear phase of evolution new phenomenon of a fast collapse of the current sheet was found.

Thus, island instability offers a simple and straightforward scenario how the process of fast and abrupt thinning of the current layer might occur and how extremely thin and intense sheets could form in collisionless plasmas. In relation to substorms, island evolution also shows two distinctly different time scales of current growth. For the equilibrium sheets with width of about ion inertia length, the period of linear and non-linear increase takes roughly hundreds ion cyclotron times. At the values $\omega_{ci} \sim 0.1-1\ s^{-1}$ typical in the Earth's magnetotail it is comparable to the duration of sluggish growth phase. The collapse of current at the neutral line has duration of about ten ion cyclotron periods and could be well associated with explosive intensification phase. When the width of intense structure becomes comparable to electron inertia length $\lambda_e$ or the current speed $J/ne$ exceeds thermal electron velocity, system goes beyond the simple Hall MHD model used here and such effects should be taken into account as electron mass, anomalous resistivity due to current-driven instabilities, kinetic interactions etc. For example, in [19] numerical simulation based on the Hall MHD with electron inertia taken into account revealed that intense thin current sheet may disrupt or disintegrate into filaments. Further investigation in the frame of a more elaborate MHD model has a promise to reveal how the described collapse may result in an extremely fast disruption, restructuring of the current sheet and onset of quasi-steady or impulsive reconnection. Also, it might provide an insight into such important and still not well understood process as particle acceleration. The other question that needs separate study is how the single islands might appear in the magnetotail. Though the perturbation of magnetic field caused by such islands could be quite small, they involve local change of topological structure.

**Acknowledgements.** Helpful discussions with Zakharov Yu.P. are appreciated.